\documentclass{svmult}

\usepackage{mathptmx}        
\usepackage{helvet}          
\usepackage{courier}         
\usepackage{type1cm}        

\usepackage{makeidx}         
\usepackage{graphicx}        
\usepackage{multicol}        
\usepackage[bottom]{footmisc}

\usepackage{amsmath,amssymb,bbm}
\usepackage{graphicx}
\usepackage[utf8]{inputenc}
\usepackage{url}
\usepackage{microtype}
\usepackage{import}
\usepackage{color}
\usepackage{epic}
\usepackage{tikz}
\usepackage{algpseudocode}
\usepackage{setspace}
\usepackage{tabularx}
\usepackage{array}

\usepackage{amsfonts}
\usepackage{psfrag}
\usepackage{epsfig}
\usepackage{multirow}
\usepackage{bm}
\usepackage{cite}
\usepackage{float}

\usepackage{mydef}

\DeclareMathAlphabet{\mathcal}{OMS}{cmsy}{m}{n}

\let\jmath=\undefined
\DeclareSymbolFont{jmathcmletters}{OML}{cmm}{m}{it}
\DeclareMathSymbol{\jmath}{\mathord}{jmathcmletters}{"7C}

\usepackage{bm}

\def\ZZZ{{Z}}

\def\BBB{{B}}

\def\NNN{{N}}

\def\WWW{{W}}

\def\ccc{{c}}

\def\iiii{{i}}

\def\nnn{{n}}

\def\yyy{{x}}

\def\psig{{\psi}}

\def\vecy{{\bm x}}

\def\Bgz{{\BBB}}

\def\cgz{{\ccc}}
\def\ggz{u^n}
\def\hgzt{{\Delta x}}
\def\hgztt{{\Delta x}}
\def\igz{{\iiii}}

\def\lgz{{L}}

\def\Wgz{{\WWW}}
\def\Zgz{{\ZZZ}}

\def\vlgz{{\ll}}
\def\vigz{{\veci}}

\def\alphgz{{\alpha}}

\def\mb{\mathbb}

\def\ll{\mathbf{l}}
\def\veci{\mathbf{i}}

\newcommand{\Seqn}{\begin{equation}}
\newcommand{\Feqn}{\end{equation}}

\newtheorem{scheme}{Scheme}

\title*{A sparse-grid probabilistic scheme for approximation of the runaway probability of electrons in fusion tokamak simulation\thanks{This material is based upon work supported in part by the U.S.~Department of Energy, Office of Science, Offices of Advanced Scientific Computing Research and Fusion Energy Science, and by the Laboratory Directed Research and Development program at the Oak Ridge National Laboratory, which is operated by UT-Battelle, LLC, for the U.S.~Department of Energy under Contract DE-AC05-00OR22725.\vspace{-0.2cm}}}

\titlerunning{A sparse-grid probabilistic scheme for runway electron simulation}

\author
{Minglei Yang\thanks{Department of Mathematics and Statistics, Auburn University,  Auburn, AL 36849, USA.\vspace{-0.2cm}}
\and
Guannan Zhang\thanks{
Computer Science and Mathematics Division,
Oak Ridge National Laboratory,
P.O.~Box 2008, MS-6211, Oak Ridge, TN 37831-6211, USA
({\tt zhangg@ornl.gov}).\vspace{-0.2cm}}
\and Diego del-Castillo-Negrete\thanks{Fusion Energy Division, Oak Ridge National Laboratory, Oak Ridge, TN 37831, USA.\vspace{-0.2cm}}
\and
Miroslav Stoyanov\thanks{
Computer Science and Mathematics Division,
Oak Ridge National Laboratory,
Oak Ridge, TN 37831-6211, USA.\vspace{-0.2cm}}
\and Matthew Beidler\thanks{Fusion Energy Division, Oak Ridge National Laboratory, Oak Ridge, TN 37831, USA.\vspace{-0.2cm}}
}

\begin{document}
\graphicspath{{figures/}} 

\numberwithin{equation}{section}
\renewcommand{\theequation}{\thesection.\arabic{equation}}

\maketitle

\newcommand{\tabincell}[2]{\begin{tabular}{@{}#1@{}}#2\end{tabular}}

\abstract{Runaway electrons (RE) generated during magnetic disruptions present a major threat to the safe operation of plasma nuclear fusion reactors. A critical aspect of understanding RE dynamics is to calculate the runaway probability, i.e., the probability that an electron in the phase space will runaway on, or before, a prescribed time. Such probability can be obtained by solving the adjoint equation of the underlying Fokker-Planck equation that controls the electron dynamics. In this effort, we present a sparse-grid probabilistic scheme for computing the runaway probability. The key ingredient of our approach is to represent the solution of the adjoint equation as a conditional expectation, such that discretizing the differential operator reduces to the approximation of a set of integrals. Adaptive sparse grid interpolation is utilized to approximate the map from the phase space to the runaway probability. 
{\color{black} The main novelties of this effort are the integration of the sparse-grid method into the probabilistic numerical scheme for computing escape probability, as well as the demonstration in computing RE probabilities. Two numerical examples are given to illustrate that the proposed method can achieve $\mathcal{O}(\Delta t)$ convergence, as well as the adaptive refinement strategy can effectively handle the sharp transition layer between the runaway and non-runaway regions.}  
}

%


\section{Introduction}\label{intro}

In magnetically confined fusion plasmas, runaway electrons (RE) can be generated during magnetic disruptions due to the strong electric field resulting from the rapid cooling of the plasma,
see for example \cite{doi:10.1063/1.4913582} and references therein.
At high enough velocities, the drag force on an electron due to Coulomb collisions in a plasma decreases as the particle velocity increases. As a result, in the presence of a strong enough parallel electric field, fast electrons can ``runaway" and be continuously accelerated, see for example the review in \cite{1979NucFu..19..785K} and references therein.  Understanding this phenomena has been an area of significant interest because of the potential impact that RE can have to the safe operation of the international test nuclear fusion reactor ITER\footnote{ITER (originally the International Thermonuclear Experimental Reactor) is an international nuclear fusion research and engineering mega project, which will be the world's largest magnetic confinement plasma physics experiment. See https://www.iter.org/ for details.}. In particular, if not avoided or mitigated, RE can severely damage plasma facing components \cite{doi:10.1063/1.4931166,doi:10.1063/1.4901251,doi:10.1063/1.3695000}. 

In this work, we propose a sparse-grid probabilistic scheme to study RE dynamics in phase space. Although the full RE model is defined in a six-dimensional phase space, here we focus on the RE dynamics in a three-dimensional space with coordinates $(p, \xi, r)$, where 
$p$ denotes the magnitude of the relativistic momentum, $\xi$ the cosine of the 
pitch angle $\theta$, i.e. the angle between the electron's velocity and the local magnetic field, and $r$ the minor radius. In this case, the dynamics 
of the distribution function of electrons is determined by the 
Fokker-Planck (FP) equation describing the competition between the electric field acceleration, Coulomb collisions, synchrotron radiation damping, and sources describing the second generation of RE due to head-on collisions \cite{Rosenbluth_1997}. 
%
%
%

In the study of RE, a set of important questions involve statistical observables different from the electron distribution function. Examples of particular interest to the present paper are 
the runaway probability, $P_{\rm RE}(t,p,\xi, r)$, that an electron with phase space coordinates $(p, \xi, r)$ will runaway on, or before, a time $t>0$. Mathematically, $P_{\rm RE}(t,p,\xi, r)$ is the solution of the adjoint of the FP equation, which is a backward parabolic equation in a non-divergence form. It is known that the non-divergence structure of the adjoint equation prevents the
use of integration by parts on it to define weak solutions, which is a pre-requisite for
formulating finite element methods for this problem. The second challenge is that the coefficients of the adjoint equation are usually very complicated, such that it is hard to convert the non-divergence operator to a divergence operator. Thus, the most widely used approach to approximate $P_{\rm RE}(t,p,\xi, r)$ is ``brute-force" Monte Carlo, which is robust and parallelizable, but features very slow convergence.


The  method we are proposing is different from those based on the solution of the Fokker-Planck equation, e.g., \cite{Marsch2006,doi:10.1063/1.4938510}, and also different from the direct Monte-Carlo simulations. 
{\color{black} Instead, our approach is based on the Feynman-Kac formula, which 
establishes a link between the adjoint of the FP equation and the system of stochastic differential equations (SDEs).}
 Specifically, we first represent the solution of the adjoint equation as a conditional expectation with respect to the underlying SDEs that describe the dynamics of the electrons. As such, the task of discretizing the differential operator becomes 
approximating the conditional expectation, which includes a quadrature rule for numerical integration and an interpolation strategy for evaluating the integrand at quadrature points. In this work, we use local hierarchical sparse grid methods\cite{Griebel1998,Bungartz2004,Klimke:2005p5580,Pfluger:2010tf,Gerstner:2003do} to handle the interpolation for two reasons. First, the terminal condition of the adjoint equation is discontinuous, and the adaptive refinement strategy of sparse grids can effectively capture such irregularity as well as well control the growth the total number of grid points. Second, the three-dimensional RE model is a simplification of the full six-dimensional model, and the use of sparse grids can make it easy to extend to the full RE model in the future work. 

{\color{black} In the literature, sparse grid methods have been applied to various plasma physics problems to approximate physical quantities in the high-dimensional phase space. For instance, sparse grids were combined with PDE solvers, e.g., discontinuous Galerkin approaches, to solve gyrokinetic equations, e.g., Vlasov-Maxwell equations \cite{TAO2019100022,10.1145/3148226.3148229}, and The Vlasov-Poisson equations \cite{10.1007/978-3-319-28262-6_7}. Not surprisingly, sparse grids were also integrated into particle-in-cell schemes \cite{Ricketson_2016} to dramatically increase the size of spatial cells and reduce the statistical noise without increasing the number of particles. In addition, scalable and resilient sparse grid techniques were applied to large-scale  
gyrokinetic problems \cite{10.1007/978-3-319-68394-2_31,10.1145/3148226.3148229,doi:10.1177/1094342015628056} to significantly accelerate existing gyro-kinetics simulators, e.g., Gyrokinetic Electromagnetic Numerical Experiment (GENE)\footnote{http://genecode.org/}. In \cite{2018arXiv181200080F}, Leja sequence based sparse interpolation has been used to analyzing gyrokinetic micro-instabilities. This effort brings another important application of the sparse grid methods to the plasma physics community. Compared to existing works in the literature, the main contribution of this effort lies in two aspects, i.e.,
\begin{itemize}\itemsep0.1cm

\item Integration of the sparse-grid method into the probabilistic numerical scheme for approximating multi-dimensional escape probability with $\mathcal{O}(\Delta t)$ convergence.

\item Demonstration of the proposed scheme in computing the probability that electrons will runaway from magnetic confinement in nuclear fusion reactors.

\end{itemize}
}


The rest of the paper is organized as follows. In Section \ref{pro_set}, we present the three-dimensional phase space runaway electron model in the  particle-based Langevin formulation, as well as its connection with the adjoint equation.  
Section~\ref{sec:sg} discusses the mathematical foundation and the numerical algorithm of the proposed method.
Section~\ref{sec:ex} represents the numerical tests of the proposed method for a two-dimensional Brownian motion problem, as well as the three-dimensional RE problem.  A summary and concluding remarks are presented in Section \ref{sec:con}.

\section{Problem setting}\label{pro_set}
We consider a three-dimensional runaway electron model describing the dynamics of the magnitude of the relativistic momentum, denoted by $p$, the cosine of the pitch angle $\theta$, denoted by $\xi=\cos \theta$, and the minor radius, denoted by $r$. The relativistic momentum $p$ is normalized using the thermal momentum and the time is normalized using the thermal collisional frequency. That is, if $\hat{p}$ and $\hat{t}$ denote the dimensional variables, then $p = \hat{p}/(mv_T)$ and $t = \nu_{ee} \hat{t}$, where $v_T = \sqrt{2T/m}$ is the thermal speed with $T$ the plasma temperature and $m$ the electron mass, and the thermal collision frequency is $\nu_{ee} = e^4 n \ln n \Lambda / (4\pi \varepsilon_0 m^2 v_T^3)$ with $e$ the absolute value of the electron charge, $\varepsilon_0$ is the vacuum permittivity and $\Lambda$ the Coulomb logarithm. The electric field is normalized using the Dreicer electric field $E_D$. Specifically, the dynamics are modeled by the following stochastic differential equations (SDEs)
\begin{equation}\label{e1}
\left\{
\begin{aligned}
dp &=  \left [E \xi\, - \frac{\gamma p}{\tau}(1-\xi^2)-C_F +\frac{1}{p^2}\frac{ \partial }{\partial p} \left(p^2 C_A\right) \right ] dt + \sqrt{2 C_A} \, dW_p,\\ 
d \xi &= \left[ \frac{E \left(1-\xi^2\right)}{p} - \frac{\xi (1-\xi^2)}{\tau \gamma}
-2 \xi \frac{C_B}{p^2} \right ] dt + \frac{\sqrt{2 C_B}}{p}  \,  \sqrt{1-\xi^2}\, dW_\xi, \,  \\
dr & =  \sqrt{2D_r} dW_r,
\end{aligned}
\right.
\end{equation}
where $W_p$, $W_\xi$ and $W_r$ are the standard Brownian motions, $E$ is the electric field, and the functions $C_A$, $C_B$, $C_F$ and $D_r$ are defined by 
\begin{eqnarray}
C_A (p) &=& \bar{\nu}_{ee} \, \bar{v}_T^2 \,\,\frac{\psi(y)}{y},  \nonumber
 \\
C_B (p)&=& \frac{1}{2} \,\bar{\nu}_{ee} 
\, \bar{v}_T^2 \, \, \frac{1}{y}  \left[ Z + \phi(y)- \psi(y) +  \frac{y^2}{2} \delta^4 \right], \nonumber\\[0.3cm]
C_F (p)&=&2\,\bar{\nu}_{ee}  \, \bar{v}_T \, \psi(y), \,  \nonumber\\[0.3cm]
D_r(p) & = & D_0 \exp(-(p/\Delta p)^2), \nonumber
\end{eqnarray}
$$
\phi(y)=\frac{2}{\sqrt{\pi}} \int_0^y e^{-s^2} ds \, ,\qquad
\psi(y)=\frac{1}{2 y^2} \left[ \phi(y)-y \frac{d \phi}{dy} \right],  \nonumber
$$
$$
y = \frac{p}{\gamma},\qquad \gamma = \sqrt{1 + (\delta p)^2}, \qquad \delta = \frac{v_T}{c} = \sqrt{\frac{2T}{mc^2}},
$$
with $Z$, $c$ denoting the ion effective charge and the speed of light, respectively.

The  problem we want to address is the computation of the probability that an electron with coordinates $(p,\xi, r)$ will runaway at, or before, a prescribed time instance.  By ``runaway" we mean that, as a result of the electric field acceleration, the electron will reach a 
prescribed momentum, $p_{\max}$. The dependence of the runaway probability on $p_{\max}$ becomes negligible for large enough $p_{\max}$, which is the reason why this dependence is not usually accounted for explicitly. 
More formally, for a given 
$(t,p, \xi, r) \in [0,T_{\max}] \times  [p_{\min}, p_{\max}] \times [-1,1] \times [0,1]$, where $p_{\min}$ is a lower momentum boundary,
the runaway probability, $P_{\rm RE}(t, p, \xi,r)$, is defined as 
the probability that an electron located at  $(p,\xi,r)$ at the initial time $t_0 = 0$ will acquire a momentum $p_{\max}$ on, or before $t >0$. 

Mathematically, the runaway probability can be described as the escape probability of a stochastic dynamical systems. For notational simplicity, we define 
\[
\bm X_t := (p, \xi, r),
\]
and rewrite the SDE in \eqref{e1} using $\bm X_t$, i.e.,
\begin{equation}\label{e2}
 \bm X_t  =   \bm X_0 + \int_0^t {\bm b}(\bm X_s) ds
+  \int_0^t{ \sigma}(\bm X_s) d  \bm W_s \;\; \text{ with } \bm X_0 \in \mathcal{D} \subset \mathbb{R}^3,
\end{equation}
where $\mathcal{D} = [p_{\min}, p_{\max}] \times [-1,1] \times [0,1]$, and the drift $\bm b$ and the diffusivity $\bm \sigma$ can be easily defined based on Eq.~\eqref{e1}. In the following sections, we will use \eqref{e2} to introduce our probabilistic scheme and will come back to Eq.~\eqref{e1} in the section of numerical examples. 

We divide the boundary of $\mathcal{D}$ into three parts $\partial \mathcal{D}_1$, $\partial \mathcal{D}_2$ and $\partial \mathcal{D}_3$, defined by
\[
\begin{aligned}
\partial \mathcal{D}_1 & := \{ p = p_{\max} \} \cap \partial \mathcal{D},\\
\partial \mathcal{D}_2 & := (\{ p = p_{\min} \} \cup \{ r = 1\}) \cap \partial \mathcal{D},\\
\partial \mathcal{D}_3 & := (\{ \xi = -1 \} \cup \{ \xi =1\} \cup \{r = 0\}) \cap \partial \mathcal{D},\\
\end{aligned}
\]
such that $\partial \mathcal{D}_1 \cup \partial \mathcal{D}_2 \cup \partial \mathcal{D}_3 = \partial \mathcal{D}$. The boundary $\partial \mathcal{D}_1$ represents the runaway boundary.
 To give a formal definition of the runaway probability, we denote the runaway time of $\bm X_t$ by
\[
\tau := \inf \big\{ t > 0 \,|\, \bm X_t \in \partial \mathcal{D}_1\big\},
\]
which represents the earliest escape time of the process $\bm X_t$ that initially starts from $\bm X_0 = \bm x \in \mathcal{D}$. Then, the runaway probability can be formally defined by
\begin{equation}\label{e2}
P_{\rm RE}(t, \bm x) = \mathbb{P}\left\{ \tau \le t \,|\, \bm X_0 = \bm x \in \mathcal{D} \right\}.
\end{equation}

For a fixed $T \in [0,T_{\max}]$, the probability $P_{\rm RE}(T, \bm x)$ can be represented by the solution of the adjoint equation of the Fokker-Planck equation based on \eqref{e2}. Such adjoint equation is a backward parabolic terminal boundary value problem, i.e.,
\begin{equation}\label{e3}
\begin{aligned}
\frac{\partial u(t,\bm x)}{\partial t}  + \mathcal{L}^{*}(t, \bm x) [u(t, \bm x)] & = 0 \quad \text{ for }\;\; \bm x \in \mathcal{D}, t < T,\\
u(t,\bm x) & = 1  \quad \text{ for }\;\; \bm x \in \mathcal{\partial D}_1, t\le T,\\[2pt]
u(t,\bm x) & = 0  \quad \text{ for }\;\; \bm x \in \mathcal{\partial D}_2, t\le T,\\[2pt]
 \nabla u(t,\bm x) & = 0  \quad \text{ for }\;\; \bm x \in \partial{\mathcal{D}}_3, t\le T,\\[2pt]
 u(T, \bm x) & = 0  \quad \text{ for }\;\; \bm x \in \mathcal{D},\\[2pt]
\end{aligned}
\end{equation}
where the operator $\mathcal{L}^{*}(t,x)$ is the adjoint of the Fokker-Planck operator, defined by
\[
\mathcal{L}^{*}(\bm x)[u] := \sum_{i=1}^{d}b_{i}\frac{\partial u}{\partial x^{i}} + \frac{1}{2}\sum_{i,j=1}^{d}({\sigma}{\sigma}^{\top})_{i,j} \frac{\partial^2 u}{\partial x^{i}x^{j}},
\]
where $b_i$ is the $i$-th component of the drift $\bm b(x)$, $({\sigma}{\sigma}^{\top})_{i,j}$ is the $(i,j)$-th entry of ${\bm \sigma}{\bm \sigma}^{\top}$ and $x^i$ is the $i$-th component of $\bm x$. It is easy to see that $P_{\rm RE}(T, \bm x)$ can be represented by 
\begin{equation}\label{e5}
P_{\rm RE}(T, \bm x) = u(0, \bm x).
\end{equation}

It should be noted that the runaway probability at each time $T$ requires a solution of the adjoint equation in \eqref{e3}, such that recovering the entire dynamics of $P_{\rm RE}$ in $[0, T_{\max}]$ requires a sequence of PDE solutions. However, due to the time {\em independence} of $\bm b$ and $\bm \sigma$ in \eqref{e2} considered in this work, the dynamics of $P_{\rm RE}(t,\bm x)$ for $(t, \bm x) \in [0,T_{\max}] \times \mathcal{D}$ can be represented by
\begin{equation}\label{e6}
P_{\rm RE}(t, \bm x) = u(T_{\max} - t, \bm x)\;\;\; \text{ for }\; t \in [0, T_{\max}],
\end{equation}
where $u$ is the solution of \eqref{e3} with $T= T_{\max}$.

\section{A sparse-grid probabilistic method for the adjoint equation}\label{sec:sg}
The theoretical foundation of our probabilistic scheme is the Feynman-Kac theory that links the SDE in Eq.~\eqref{e2} to the adjoint problem in Eq.~\eqref{e3}. This section focuses on solving the adjoint equation in Eq.~\eqref{e3}. 
The probabilistic representation of $v(t,x)$ and the temporal discretization is given in Section \ref{sec:time}; spatial discretization including a special treatment of the involved random escape time $\tau$ is provided in Section \ref{sec:exp} and \ref{sec:interp}.

\subsection{Temporal discretization}\label{sec:time}
 To write out the probabilistic representation of $u(t,x)$ in Eq.~\eqref{e3}, we need to rewrite the SDE in Eq.~\eqref{e2} in a conditional form, i.e.,
\begin{equation}\label{eq:SDE}
  \bm X_s^{t,  x}  =   \bm x + \int_t^s { \bm b}(\bm X_{\bar{t}}^{t,  x}  ) d\bar{t}
+  \int_t^s{\bm   \sigma}(\bm X_{\bar{t}}^{t,  x}  ) d \bm W_{\bar{t}} \;\; \text{ for } \; s \ge t,
\end{equation}
where the superscript ${}^{t, x}$ indicates the condition that $\bm X_s^{t,x}$ starts from $(t,\bm x) \in [0,T_{\max}]  \times \mathcal{D}$. Accordingly, we can define the \emph{conditional escape time} \cite{oksendalSDE}
\begin{equation}\label{e22}
\tau_{t,   x}  := \min(\tau_{t,x}^1, \tau_{t,x}^2)
\end{equation}
with
\begin{equation}\label{e29}
\tau_{t,   x}^1  :=
\inf\{ s > t \,|\,   \bm X_s^{t,   x}\in \partial \mathcal{D}_1\}, \;\; \tau_{t,   x}^2 :=
\inf\{ s > t \,|\,   \bm X_s^{t,   x}\in \partial \mathcal{D}_2\},
\end{equation}
such that the probabilistic representation of the solution $u(t, \bm x)$ of the adjoint equation in \eqref{e3} is given in \cite{Peng:1990vu,Pardoux:1998wg}, i.e., 
\begin{equation}\label{e23}
u(t,\bm x) = \mathbb{E}\left[u \left(s\wedge \tau_{t,x}, \bm X_{s\wedge \tau_{t,x}}^{t,  x} \right) \right],
\end{equation}
where $s \wedge \tau_{t,x}$ denotes the minimum of $\tau_{t,x}$ and $s$, $\tau_{t,x}$ is given in Eq.~\eqref{e22}, and $\bm X^{t,x}_{s \wedge \tau_{t,x}}$ is defined based on Eq.~\eqref{eq:SDE}. 

We then discretize the probabilistic representation of $u$ in Eq.~\eqref{e23}. To proceed, we introduce a uniform time partition for $[0,T_{\max}]$:
\[
\mathcal{T} := \{0=t_0<t_1<\cdots<t_{N}=T_{\max}\}
\]
with $\Delta t=t_{n+1}-t_n$ and $\Delta \bm W := \bm W_{t_{n+1}}- \bm W_{t_n}$ for $n = 0,1, \ldots, N$. The SDE in Eq.~\eqref{e2} can be discretized in the interval $[t_n, t_{n+1}]$ using the forward Euler scheme:
\begin{equation}\label{ref-X}
\bm X_{n+1}^{t_n,x} = \bm x + \bm b(\bm x)\Delta t+ \bm \sigma(\bm x)\Delta \bm W, 
\end{equation}
such that the Eq.~\eqref{e23} can be discretized (see \cite{Yang:2018fd}) as
\begin{equation}\label{e24}
u^n(\bm x) = \mathbb{E}\left[ u^{n+1}\left(\bm X_{n+1}^{t_n, x}\right)\mathbf{1}_{\{\tau_{t_n,x} > t_{n+1}\}}\right] +  \mathbb{P}\left(\tau_{t_n,x}^1 \le t_{n+1}\right),
\end{equation}
where $u^n(x) \approx u(t_n,x)$, $\tau_{t_n, x}$ is defined in Eq.~\eqref{e22}\footnote{The escape time $\tau_{t_n,x}$ in Eq.~\eqref{e24} should be defined by replacing $\bm X_s^{t,x}$ with the Euler discretization, i.e., $\bm X_{s}^{t_n,x} = \bm x + b(\bm x)(s-t_n)+ \sigma(\bm x)(\bm W_s-\bm W_{t_n})$ for $s\ge t_n$ in Eq.~\eqref{e22}. We use the same notation without creating confusion.},
and $\mathbf{1}_{\{\tau_{t_n,x} > t_{n+1}\}}$ is the characteristic function of the event that $\bm X_{s}^{t_n,x}$ does not escape the domain $\mathcal{D}$ via $\partial \mathcal{D}_1 \cup \partial \mathcal{D}_2$ before $t_{n+1}$. 

\subsection{Sparse-grid interpolation for spatial discretization}\label{sec:interp}
To extend the time-stepping scheme in Eq.~\eqref{e24} to a fully-discrete scheme, we need to a spatial discretization scheme to approximate $u^{n}$ as well as a quadrature rule to estimate the conditional expectation $\mathbb{E}[\cdot ]$. In this work, we intend to use piecewise sparse grid interpolation to approximate $u^n(x)$ in $\mathcal{D}$. Specifically, since the terminal condition of the adjoint equation in Eq.~\eqref{e3} is discontinuous, we used hierarchical sparse grids with piecewise polynomials\cite{Bungartz2004,Pfluger:2010tf}, which is easy to incorporate adaptivity to handle the discontinuity.

\subsubsection{Hierarchical sparse grid interpolation}
We briefly recall the standard hierarchical sparse grid interpolation by borrowing the main notations and results from \cite{Bungartz2004,Pfluger:2010tf,dirk,Bungartz:1996vo}. 
The one-dimensional hat function having support $[-1,1]$ is defined by
$
\psig(\yyy) = \max\{ 0\,,\,1-|\yyy| \}
$ 
 from which an arbitrary hat function with support $(\yyy_{\lgz,\igz} - \hgztt_\lgz, \yyy_{\lgz,\igz} + \hgztt_\lgz)$ can be generated by dilation and translation, i.e.,
$$
\psig_{\lgz,\igz}(\yyy) := \psig\Big(\frac{\yyy +1 - \igz  \hgztt_\lgz}{ \hgztt_\lgz}\Big),
$$ 
where $\lgz$ denotes the resolution level, $\hgztt_\lgz = 2^{-\lgz+1}$ for $\lgz = 0,1, \ldots$ denotes the grid size of the level $\lgz$ grid for the interval $[-1,1]$, and $\yyy_{\lgz,\igz} = \igz \, \hgztt_\lgz -1$ for $\igz = 0, 1, \ldots, 2^\lgz$ denotes the grid points of that grid. The basis function $\psig_{\lgz,\igz}(\yyy)$ has local support and is centered at the grid point $\yyy_{\lgz,\igz}$; the number of grid points in the level $\lgz$ grid is $2^\lgz+1$. 

One can generalize the piecewise linear hierarchical polynomials to high-order hierarchical polynomials, as shown in \cite{Bungartz2004, Bungartz:1996vo}. 
\begin{figure}[h!]
\begin{center}
\includegraphics[scale = 0.13]{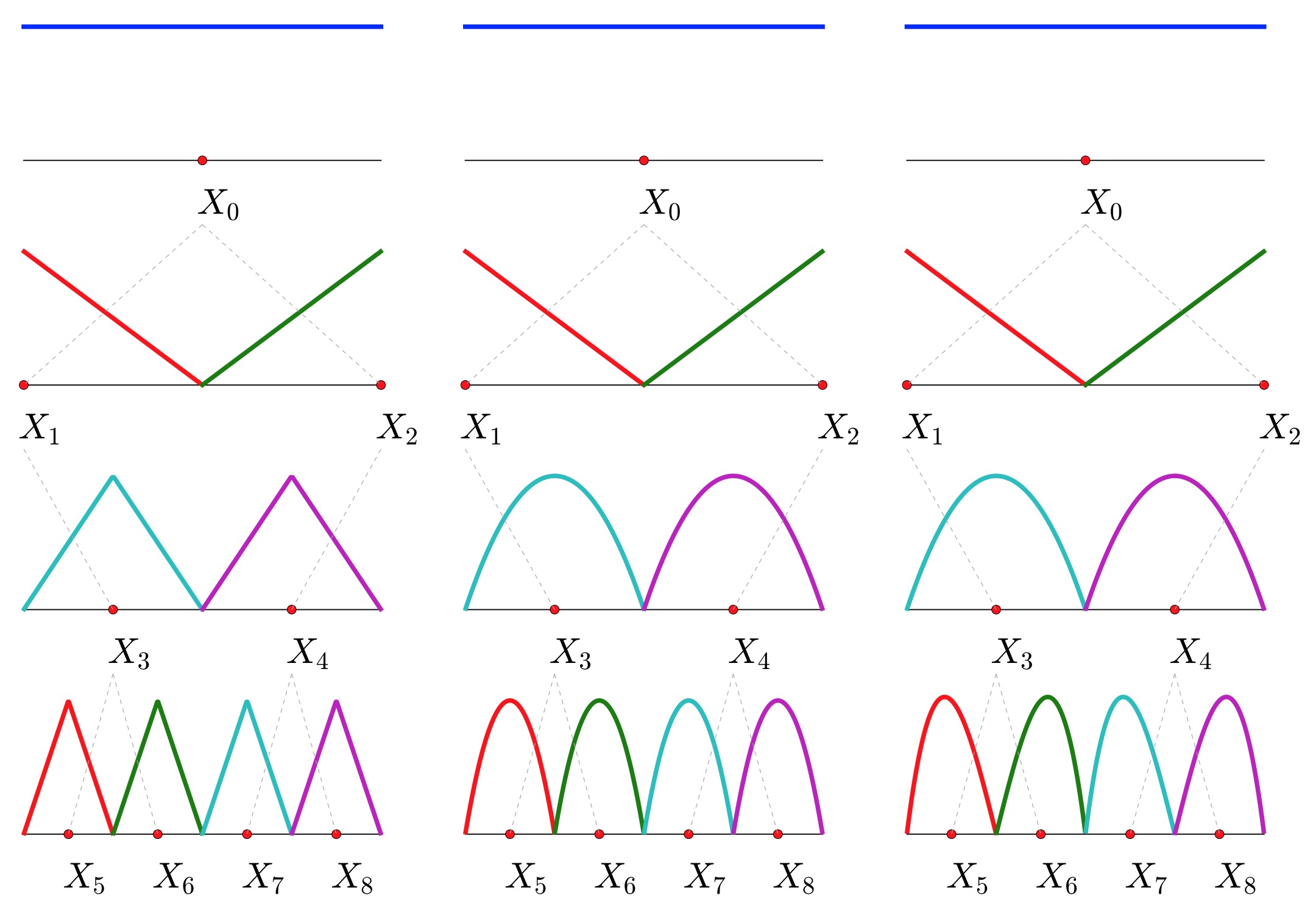}
\caption{Left: linear hierarchical basis; Middle: quartic hierarchical basis where the quadratic polynomials appear since level 2; Right: cubic hierarchical basis where the cubic polynomials appear since level 3.}\label{hier_1D2}
\end{center}
\end{figure}
As shown in Fig \ref{hier_1D2}, for $\lgz \ge 0$, a piecewise linear polynomial $\psig_{\lgz,\igz}(\yyy)$ is defined based on 3 supporting points, i.e., $\yyy_{\lgz,\igz}$ and its two ancestors that are also the endpoints of the support $[\yyy_{\lgz,\igz} - \hgzt_\lgz, \yyy_{\lgz,\igz} + \hgzt_\lgz]$. For $q$-th order polynomials, $q+1$ supporting points are needed to define a Lagrange interpolating polynomial. To do this, at each grid point $\yyy_{\lgz,\igz}$, additional ancestors outside of $[\yyy_{\lgz,\igz} - \hgzt_\lgz, \yyy_{\lgz,\igz} + \hgzt_\lgz]$ are borrowed to help build a higher-order Lagrange polynomial, then, the desired high-order polynomial is defined by restricting the resulting polynomial to 
the support $[\yyy_{\lgz,\igz} - \hgzt_\lgz, \yyy_{\lgz,\igz} + \hgzt_\lgz]$. Fig \ref{hier_1D2} illustrates the linear, quadratic and cubic hierarchical bases, respectively.

With $\Zgz = L^2(\mathcal{D})$, a sequence of subspaces $\{\Zgz_{\lgz}\}_{{\lgz}=0}^{\infty}$ of $\Zgz$ of increasing dimension $2^\lgz+1$ can be defined as
$$
\Zgz_\lgz = \text{span} \big\{\psig_{\lgz,\igz}(\yyy)\,\, |\,\,  \igz= 0,1,\ldots,2^\lgz \big\} \quad\mbox{for $\lgz = 0, 1, \ldots$}. 
$$
%
Due to the nesting property of $\{\Zgz_\lgz\}_{l=0}^{\infty}$, we can define a sequence of hierarchical subspaces as
$
\Wgz_\lgz = \text{span}\big\{ \psig_{\lgz,\igz}(\yyy) \,\,|\,\, \igz \in \Bgz_\lgz \big\}
$
where 
$
  \Bgz_{\lgz} = \big\{ \igz \in \mb{N} \,\,\,\big|\,\,\,
   \igz = 1, 3, 5, \ldots , 2^\lgz-1 \big\}$ for $\lgz=1,2,\ldots$, such that $\Zgz_\lgz = \Zgz_{\lgz-1} \oplus \Wgz_\lgz$ and $\Wgz_\lgz = \Zgz_\lgz / \oplus_{\lgz'=0}^{\lgz-1} \Zgz_{\lgz'}$ for  $\lgz=1,2,\ldots$. Then, the hierarchical subspace splitting of $\Zgz_\lgz$ is given by
$$
\Zgz_\lgz= \Zgz_0 \oplus \Wgz_1 \oplus \cdots \oplus \Wgz_\lgz 
\quad\mbox{for $\lgz=1,2,\ldots$}.
$$

%
The one-dimensional hierarchical polynomial basis can be extended to the $\NNN$-dimensional domain using sparse tensorization. Specifically, the $\NNN$-variate basis function $\psig_{\vlgz,\vigz}(\vecy)$ associated with the point $\vecy_{\vlgz,\vigz} = (\yyy_{\lgz_1,\igz_1}, \ldots, \yyy_{\lgz_\NNN, \igz_\NNN})$ is defined using tensor products, i.e.,
$
\psig_{\vlgz,\vigz}(\vecy) := \prod_{n=1}^\NNN \psig_{\lgz_\nnn, \igz_\nnn} (\yyy_\nnn),
$
where $\{\psig_{\lgz_\nnn, \igz_\nnn} (\yyy_\nnn)\}_{\nnn=1}^\NNN$ are the one-dimensional hierarchical polynomials associated with the point $\yyy_{\lgz_\nnn, \igz_\nnn} = \igz_\nnn  \hgzt_{\lgz_\nnn}-1$ with $\hgzt_{\lgz_\nnn} = 2^{-\lgz_n+1}$ and  $\vlgz = (\lgz_1, \ldots, \lgz_\NNN)$ is a multi-index indicating the resolution level of the basis function. The $\NNN$-dimensional hierarchical incremental subspace $\Wgz_{\vlgz}$ is defined by
$$
\Wgz_{\vlgz} = \bigotimes_{\nnn=1}^\NNN \Wgz_{\lgz_\nnn} = \text{span} 
\left\{ \left. \psig_{\vlgz,\vigz}(\vecy) \,\right|\, \vigz \in B_{\vlgz}\right\},
$$
where the multi-index set $\Bgz_{\vlgz}$ is given by
$$
  \Bgz_{\vlgz} := \left\{  \vigz \in \mb{N}^{\NNN} \;\Bigg|\;
  \begin{aligned}
 & \mbox{$\igz_\nnn \in\{1,3,5,\ldots, 2^{\lgz_\nnn}-1\}$} & \mbox{for $\nnn=1,\ldots,\NNN$} & \quad\mbox{if $\lgz_\nnn > 0$} \\
 & \igz_\nnn \in\{ 0,1\} & \mbox{for $\nnn=1,\ldots,\NNN$} & \quad\mbox{if $\lgz_\nnn  = 0$}
\end{aligned}
\right\}.
$$
Similar to the one-dimensional case, a sequence of subspaces, again denoted by $\{\Zgz_\lgz\}_{\lgz=0}^{\infty}$, of the space $\Zgz := L^2(\mathcal{D})$ can be constructed as
$$
 \Zgz_\lgz = \bigoplus_{\lgz'=0}^\lgz \Wgz_{\lgz'} = \bigoplus_{\lgz'=0}^{\lgz} \bigoplus_{\alphgz(\vlgz') = \lgz'} \Wgz_{\vlgz'},
$$
where the key is how the mapping $\alphgz(\vlgz)$ is defined because it defines the incremental subspaces $\Wgz_{\lgz'}=\oplus_{\alphgz(\vlgz') = \lgz'} \Wgz_{\vlgz'}$. For example, $\alphgz(\vlgz) = |\vlgz| = \lgz_1 + \ldots + \lgz_\NNN$ leads to a standard isotropic sparse polynomial space.

The level $\lgz$ hierarchal sparse grid interpolant of the approximation $u^{n}(\bm x)$ in Eq.~\eqref{e24} is defined by
\begin{equation}\label{SGinterp}
\begin{aligned}
\ggz_\lgz(\vecy) & := \sum_{\lgz'=0}^\lgz \sum_{|\vlgz'| = \lgz'} (\Delta_{\lgz'_1} \otimes \cdots \otimes \Delta_{\lgz'_\NNN}) \ggz(\vecy)\\
    & = \ggz_{\lgz-1}(\vecy) + \sum_{|\vlgz'| = \lgz} (\Delta_{\lgz'_1} \otimes \cdots \otimes \Delta_{\lgz'_\NNN}) \ggz(\vecy)\\
    & = \ggz_{\lgz-1}(\vecy) + \sum_{|\vlgz'| = \lgz} \sum_{\vigz \in \Bgz_{\vlgz'}} 
                                 \big[\ggz(\vecy_{\vlgz',\vigz}) - \ggz_{\lgz'-1}(\vecy_{\vlgz',\vigz})\big] \psig_{\vlgz',\vigz}(\vecy)\\
    & = \ggz_{\lgz-1}(\vecy) + \sum_{|\vlgz'| = \lgz} \sum_{\vigz \in \Bgz_{\vlgz'}} \cgz_{\vlgz',\vigz} \, \psig_{\vlgz',\vigz}(\vecy),
\end{aligned}
\end{equation}
where $\ccc_{\vlgz',\vigz} =\ggz(\vecy_{\vlgz',\vigz}) - \ggz_{\lgz'-1}(\vecy_{\vlgz',\vigz})$ is the multi-dimensional hierarchical surplus \cite{10.1145/1114268.1114275}. This interpolant is a direct extension, via the Smolyak algorithm\cite{Smolyak_63}, of the one-dimensional hierarchical interpolant.

\subsubsection{A strategy for handling the boundary condition}\label{sec:bound}
After the sparse grid, denoted by $\mathcal{S}$, is constructed, the task becomes to estimate the right-hand side of Eq.~\eqref{e24} at all the interior sparse grid points $\bm x_\ii \in \mathcal{S} \cap \mathcal{D}$. The accuracy of such estimation also depends on how to deal with $\mathbb{P}(\tau_{t_n,x}^1 \le t_{n+1})$. It is known that $\mathbb{P}(\tau_{t_n,x}^1 \le t_{n+1}) \rightarrow 1$ as $\bm x \rightarrow \partial \mathcal{D}_1$. In our previous work \cite{Yang:2018fd}, we proved, in the one-dimensional case, that if $b$ and $\sigma$ are bounded functions, i.e.,
\[
|b(t,x)| \le \overline{b} \;\; \text{ and } \;\; |\sigma(t,x)| \le \overline{\sigma} \;\; \text{ for } \;\; (t,x) \in [0,T]\times \mathcal{D},
\] 
with $0 \le \overline{b}, \overline{\sigma} \le +\infty$, and the starting point $x$ in Eq.~\eqref{ref-X} is sufficiently far from the boundary $\partial \mathcal{D}$ satisfying 
$
dist(x, \partial \mathcal{D}) \sim \mathcal{O}((\Delta t)^{1/2-\varepsilon})
$
 for any given constant $\varepsilon >0$, then for sufficiently small $\Delta t$, it holds that
\begin{equation}\label{stoperr}
\mathbb{P}(\tau_{t_n,x}^1 \le t_{n+1}) \le C (\Delta t)^\varepsilon \exp\left(-\frac{1}{(\Delta t)^{2\varepsilon}}\right),
\end{equation}
where the constant $C>0$ is independent of $\Delta t$.

{\color{black} Even though the estimate in Eq.~\eqref{stoperr} was proved for the one-dimensional case, we exploited the estimate in the three-dimensional to design our numerical scheme.} The key idea is to eliminate the destructive effect of $\mathbb{P}(\tau_{t_n,x}^1 \le t_{n+1})$ in the construction of the temporal-spatial discretization scheme by exploiting the estimate in Eq.~\eqref{stoperr}. Specifically, we define the spatial mesh size $\Delta x$ of the sparse grid is on the order of 
\[
\Delta x \sim \mathcal{O}\left((\Delta t)^{\frac{1}{2}-\varepsilon}\right),
\]
such that, for each interior grid point $\bm x_\ii$, $u^n(\bm x_\ii)$ in Eq.~\eqref{e24} can be approximated by
\begin{equation}\label{e30}
u^n(x_\ii) \approx \mathbb{E}\left[ u^{n+1}_{L}\left(\bm X_{n+1}^{t_n, x_\ii}\right)\right],
\end{equation}
with the error on the order of $\mathcal{O}((\Delta t)^\varepsilon \exp(-{1}/{(\Delta t)^{2\varepsilon}}))$. The specific choice of $\Delta x$ will be given in Section \ref{sec:exp}. Such strategy can avoid the approximation of the escape probability $\mathbb{P}(\tau_{t_n,x}^1 \le t_{n+1})$, but the trade-off is that we need to use higher order sparse grid interpolation to balance the total error.

\subsection{Quadrature for the conditional expectation}\label{sec:exp}
The last piece of the puzzle is a quadrature rule for estimating the conditional expectations $\mathbb{E}\left[ u^{n+1}_{L}\left(\bm X_{n+1}^{t_n,\bm x_\ii}\right)\right]$ for $\bm x_\ii \in \mathcal{S} \cap \mathcal{D}$.
Such expectation can be written as
\begin{equation}
\mathbb{E}\left[u^{n+1}_{L}(\bm X_{n+1}^{t_n, \bm x_\ii})\right] = 
\int_{\mathbb{R}^d} u^{n+1}_{L}\left(\bm x_\ii+ b(\bm x_\ii) \Delta t + \sigma(\bm x_\ii)\sqrt{2\Delta t}\, \eta \right) \rho(\eta) d\eta,
\end{equation}
where $\eta := (\eta_1, \ldots, \eta_d)$ follows the $d$-dimensional standard normal distribution, i.e.,
$
\rho(\eta) := \frac{1}{\pi^{d/2} } \exp(-\sum_{\ell=1}^d \eta_\ell^2).
$
Thus, we utilized tensor-product Gauss-Hermite quadrature rule to approximate the expectation.
Specifically, we denote by $\{w_j\}_{j=1}^J$ and $\{a_j\}_{j=1}^J$ the weights and abscissae of the $J$-point tensor-product Gauss-Hermite rule, respectively.   
Then the approximation, denoted by $\widehat{\mathbb{E}}[u^{n+1}_{L}(\bm X_{n+1}^{t_n, \bm x_\ii})]$ is defined by
%
\begin{equation}\label{quad}
u^n_\ii = \widehat{\mathbb{E}}\left[u^{n+1}_{L}(\bm X_{n+1}^{t_n,\bm x_\ii})\right] = \sum_{j = 1}^J w_{j} \; u^{n+1}_{L}(\bm q_{\ii j}),
\end{equation}
with
\begin{equation}\label{e31}
\bm q_{\ii j} := \bm x_\ii+ b(\bm x_\ii) \Delta t + \sigma(\bm x_\ii ) \sqrt{2\Delta t}\,a_{j}
\end{equation}
where $\omega_{j}$ is a product of the weights of the one-dimensional rule and $a_j$ is a $d$-dimensional vector consisting of one-dimensional abscissae, respectively. When $u^{n+1}_{L}(\cdot)$ is sufficiently smooth, i.e., $\partial^{2J^*} u^{n+1}/\partial \eta_\ell^{2J^*}$ is bounded for $\ell = 1, \ldots, d$ with $J^* = J^{1/d}$, then the quadrature error can be bounded by \cite{2013JSV...332.4403B}
\[
\left|\widehat{\mathbb{E}}[u^{n+1}_{L}(\bm X_{n+1}^{t_n,\bm x_\ii})] - {\mathbb{E}}[u^{n+1}_{L}(\bm X_{n+1}^{t_n,\bm x_\ii})]\right| \le C\frac{J^*!}{2^{J^*}(2J^{*})!} (\Delta t)^{J^*},
\]  
where the constant $C$ is independent of $J^*$ and $\Delta t$. Note that the factor $(\Delta t)^{J^*}$ comes from the $2{J^*}$-th order differentiation 
of the function $u^{n+1}$ with respect to $\eta_\ell$ for $\ell = 1, \ldots, d$. Thus, to achieve first order global convergence rate $\mathcal{O}(\Delta t)$, we only need to use a total of $J^* = 27$ quadrature points. Sparse-grid Gauss-Hermite rule could be used to replace the tensor product rule when the dimension $d$ is higher than 3. For the 3D runaway electron problem under consideration, we found that a level 1 sparse Gauss-Hermite rule with 7 quadrature points cannot provide sufficient accuracy, and a level 2 rule with 37 points is more expensive than the tensor product rule. Thus, we chose to use the tensor-product rule in this work.

%

By putting together all the components introduced in Section \ref{sec:sg}, we summarize our probabilistic scheme as follows:
\begin{scheme}[The fully-discrete probabilistic scheme]\label{s4:full}
Given the temporal-spatial partition $\mathcal{T}\times \mathcal{S}$, the terminal condition $u^N(\bm x_\ii)$ for $\bm x_\ii \in \mathcal{S}$, and the boundary condition $u^n(\bm x_\ii)$ for $\bm x_\ii \in \mathcal{S}\cap\partial {\mathcal{D}}$. For $n = N-1, \ldots, 0$, the approximation of $u(t_n, \bm x)$ is constructed via the following steps:
\begin{itemize}\itemsep0.2cm
\item Step 1: generate quadrature abscissae $\{\bm q_{\ii j}\}_{j=1}^J$, in Eq.~\eqref{e31}, for $\bm x_\ii \in \mathcal{S} \cap \mathcal{D}$;
\item Step 2: interpolate $u^{n+1}_{L}(x)$ at the quadrature abscissae to obtain $\{u^{n+1}_{L}(\rm q_{\ii j})\}_{j=1}^J$;
%
\item Step 3: compute the coefficients $u_\ii^n$ using the quadrature rule in Eq.~\eqref{quad};
\item Step 4: construct the interpolant $u^{n}_{L}(\rm x)$ by substituting $u_n^\ii$ into Eq.~\eqref{SGinterp}.
\end{itemize}

\end{scheme}

{\color{black} A major novelty of the proposed method is that Scheme \ref{s4:full} is the first numerical scheme, which integrates sparse grids into a probabilistic scheme, for computing escape probabilities of stochastic dynamical systems.
Even though the escape probability can be computed by solving the adjoint equation in Eq.~\eqref{e3} with sparse-grid-based PDE solvers, there are several significant advantages of combining sparse grids with scheme probabilistic scheme. }
First, the time-stepping scheme is fully explicit but absolutely stable, which has been rigorously proved in our previous work, e.g., \cite{Zhang:2017jn,Zhao:2010ik}. Second, the Feynman-Kac formula makes it natural to incorporate any sparse grid interpolation strategies to approximate the solution $u$ without worrying about the discretization of the differential operator on the sparse grid. Third, it is easy to incorporate legacy codes for Monte Carlo based RE simulation into our scheme to compute runaway probability. This is a valuable feature because real-world RE models usually involve complex multiscale dynamics that is challenging to solve using PDE approaches.

\section{Numerical examples}\label{sec:ex}
We tested our probabilistic scheme with two examples. In the first example, we compute the escape probability of the standard Brownian motion. Since we know the analytical expression of the escape probability, this example is used to demonstrate the accuracy of our approach. In the second example, we to compute the runaway probability of the three-dimensional RE model given in Section \ref{pro_set}. The sparse grid interpolation and adaptive refinement are implemented using the TASMANIAN toolbox \cite{stoyanov2015tasmanian}.

\subsection{Example 1: escape probability of a Brownian motion}\label{ex:BM}
We consider the escape probability of a two-dimensional Brownian motion \cite{Schuss:2013th}. The spatial domain $\mathcal{D}$ is set to $[0,5] \times [0,5]$ and the temporal domain is set to $t \in [0,2]$ with $T_{\max} = 2$. The escape probability $P(t, \bm x)$ can be obtained by solving the standard heat equation
\begin{equation}\label{ex2d}
\begin{aligned}
\frac{\partial u}{\partial t}  + \frac{1}{2}\Delta u & = 0, \quad  (t,\bm x) \in [0,T_{\max}] \times \mathcal{D},\\
u(t,\bm x) & =1 ,\quad (t, \bm x) \in  [0,T_{\max}] \times \partial \mathcal{D},\\
u(T_{\max},\bm x) & =0 ,\quad \bm x \in \mathcal{D}.
\end{aligned}
\end{equation}
The exact solution is given by
\begin{equation*}
    u(t,\bm x)=1+\sum_{m=1}^{\infty}\sum_{n=1}^{\infty}A_{mn}\sin{(\mu_{m}x_1)}\sin{(\nu_n x_2)}e^{-\lambda^2_{mn}t},
\end{equation*}
where $\mu_{m}=\frac{m\pi}{5}$, $\nu_{n}=\frac{n\pi}{5}$, $\lambda =\sqrt{\frac{1}{2(\mu_{m}^2+\nu_{n}^2)}}$. The escape probability $P(t, \bm x)$ can be obtained by substituting $u$ into Eq.~\eqref{e6}, i.e., $P(t,\bm x) = u(T_{\max}-t, \bm x)$. 

\begin{figure}[h!]
\center
  \includegraphics[scale = 0.22]{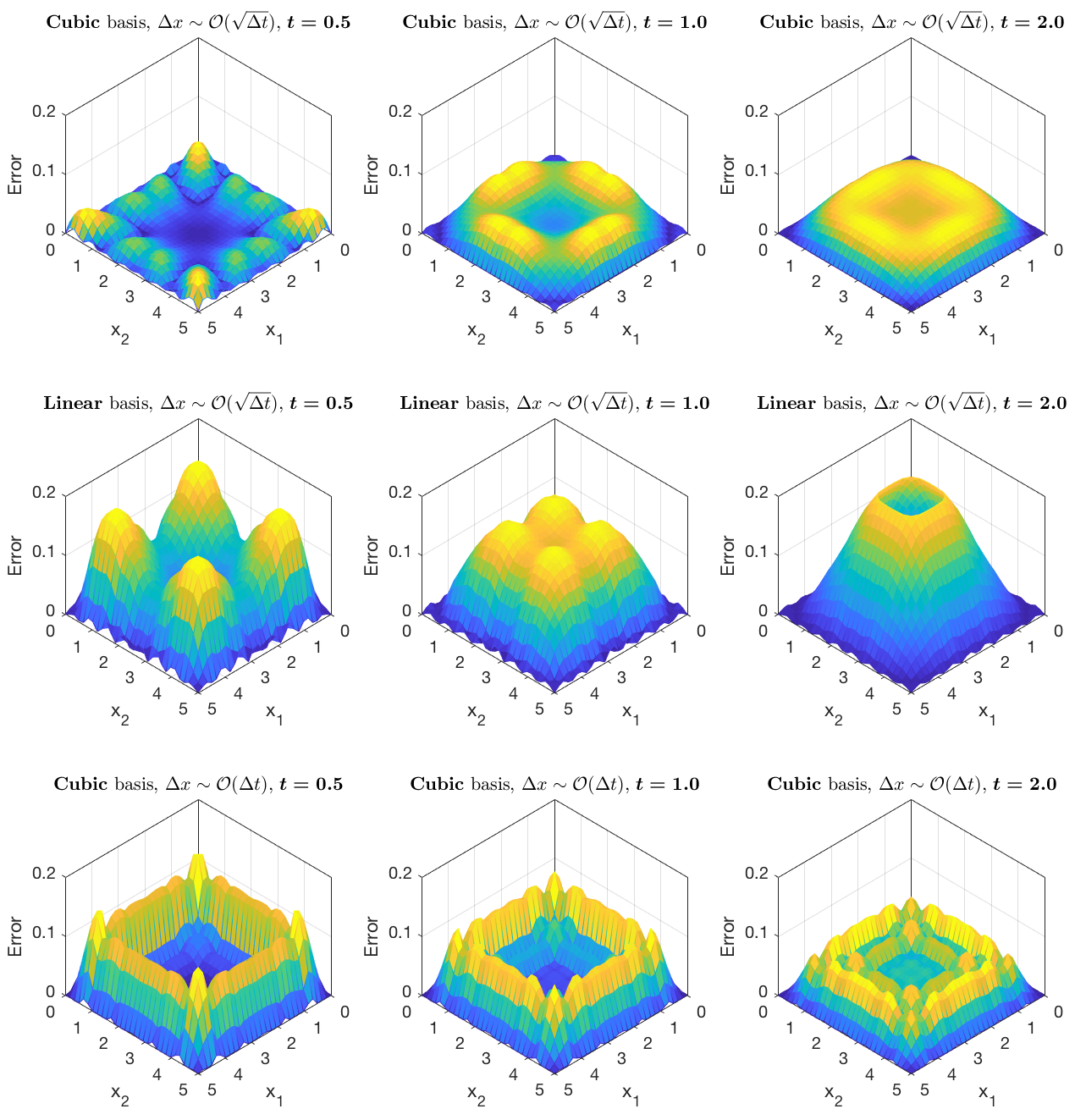}
  \caption{The error distribution in the spatial domain $[0,5]\times[0,5]$ for $t = 0.5, 1.0$ and $2.0$. The first row corresponds to the case (a), the second row corresponds to the case (b), and the third row corresponds to the case (c) in Fig \label{fig:ex2con}.}
  \label{fig:ex12}
\end{figure}

We intend to demonstrate that our scheme can achieve first-order convergence $\mathcal{O}(\Delta t)$ when properly choosing the sparse grid resolution, i.e., the level $L$. To this end, we compare three cases, i.e., 
\begin{itemize}\itemsep0.1cm
\item[(a)] Hierarchical cubic basis with $\Delta x \sim \mathcal{O}(\sqrt{\Delta t})$,
\item[(b)] Hierarchical linear basis with $\Delta x \sim \mathcal{O}(\sqrt{\Delta t})$,
\item[(c)] Hierarchical cubic basis with $\Delta x \sim \mathcal{O}(\Delta t)$,
\end{itemize}
where $\Delta x$ denotes the mesh size of the one-dimensional rule for building the sparse grids. 
The error of the three cases are shown in 
Fig \ref{fig:ex2con} and Fig \ref{fig:ex12}. As expected, in case (a), i.e., $\Delta x \sim \mathcal{O}(\sqrt{\Delta t})$, the escape probability $\mathbb{P}\left(\tau_{t_n,x} \le t_{n+1}\right)$ for any interior grid point is on the order of $\mathcal{O}((\Delta t)^\varepsilon \exp(-{1}/{(\Delta t)^{2\varepsilon}}))$, such that neglecting $\mathbb{P}\left(\tau_{t_n,x} \le t_{n+1}\right)$ will asymptotically not affect the first-order convergence w.r.t.~$\Delta t$. On the other hand, we need to use high-order hierarchical basis to achieve comparable accuracy in spatial approximation. It is shown in Fig \ref{fig:ex2con} that the use of the hierarchical cubic polynomials, introduced in \cite{Bungartz2004}, provides sufficient accuracy to achieve a global convergence rate $\mathcal{O} (\Delta t)$. In comparison, in case (b), i.e., using linear basis with $\Delta x \sim \mathcal{O}(\sqrt{\Delta t})$, the linear sparse-grid interpolation only provides $\mathcal{O}((\Delta x)^2) = \mathcal{O}(\Delta t)$ local convergence, such that our scheme dose not converge globally. From the second row of Fig \ref{fig:ex12}, we can see that large errors are generated around the boundary of the spatial domain and gradually propagate to the middle region of the domain. Similar phenomenon appears in the case (c) when setting $\Delta x \sim \mathcal{O}(\Delta t)$. In this case, the interior grid points near the boundary are so close to the boundary that neglecting the escape probability $\mathbb{P}\left(\tau_{t_n,x} \le t_{n+1}\right)$ leads to significant additional error. This is the reason why big errors are firstly generated near the boundary (i.e., $t = 0.5$), and then propagate to the center.
\begin{figure}[h!]
\center
  \includegraphics[scale=0.17]{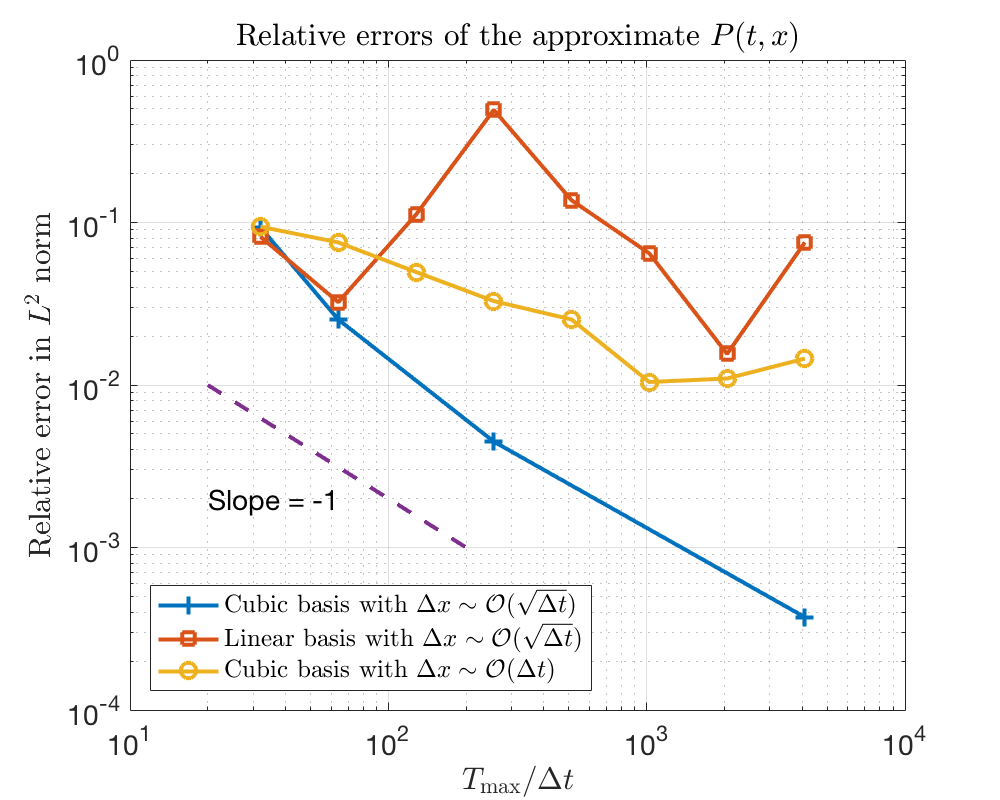}
  \caption{The relative error of the approximate escape probability of the standard Brownian motion for the three test cases, i.e., 
  (a) cubic basis with $\Delta x \sim \mathcal{O}(\sqrt{\Delta t})$, (b) linear basis with $\Delta x \sim \mathcal{O}(\sqrt{\Delta t})$, (c) cubic basis with $\Delta x \sim \mathcal{O}(\Delta t)$.}
  \label{fig:ex2con}
\end{figure}

\subsection{The runaway probability of the three-dimensional RE model}\label{sec:RE}
Here we test our method using the 3D runaway electron model given in Eq.~\eqref{e1} with the following parameters:
\[
\begin{aligned}
T_{\max} & = 120, \;\; p_{\min} = 2, \;\; p_{\max} = 50,\;\; Z = 1,\;\;  \tau = 10^5,\;\;  \delta = 0.042,\\
& E = 0.3, \;\; \bar{v}_{ee} = 1, \;\; \bar{v}_T =1,\;\; D_0 = 0.003,\;\; \Delta p = 20. 
\end{aligned}
\]
Unlike the example about Brownian motion, where the discontinuous terminal condition is smoothed out very fast, the evolution of the runaway probability $P_{\rm RE}$ is more convection-dominated. As such, we utilized adaptive sparse grids to capture the movement of the sharp transition layer. 
\begin{figure}[h!]
\center
  \includegraphics[scale = 0.22]{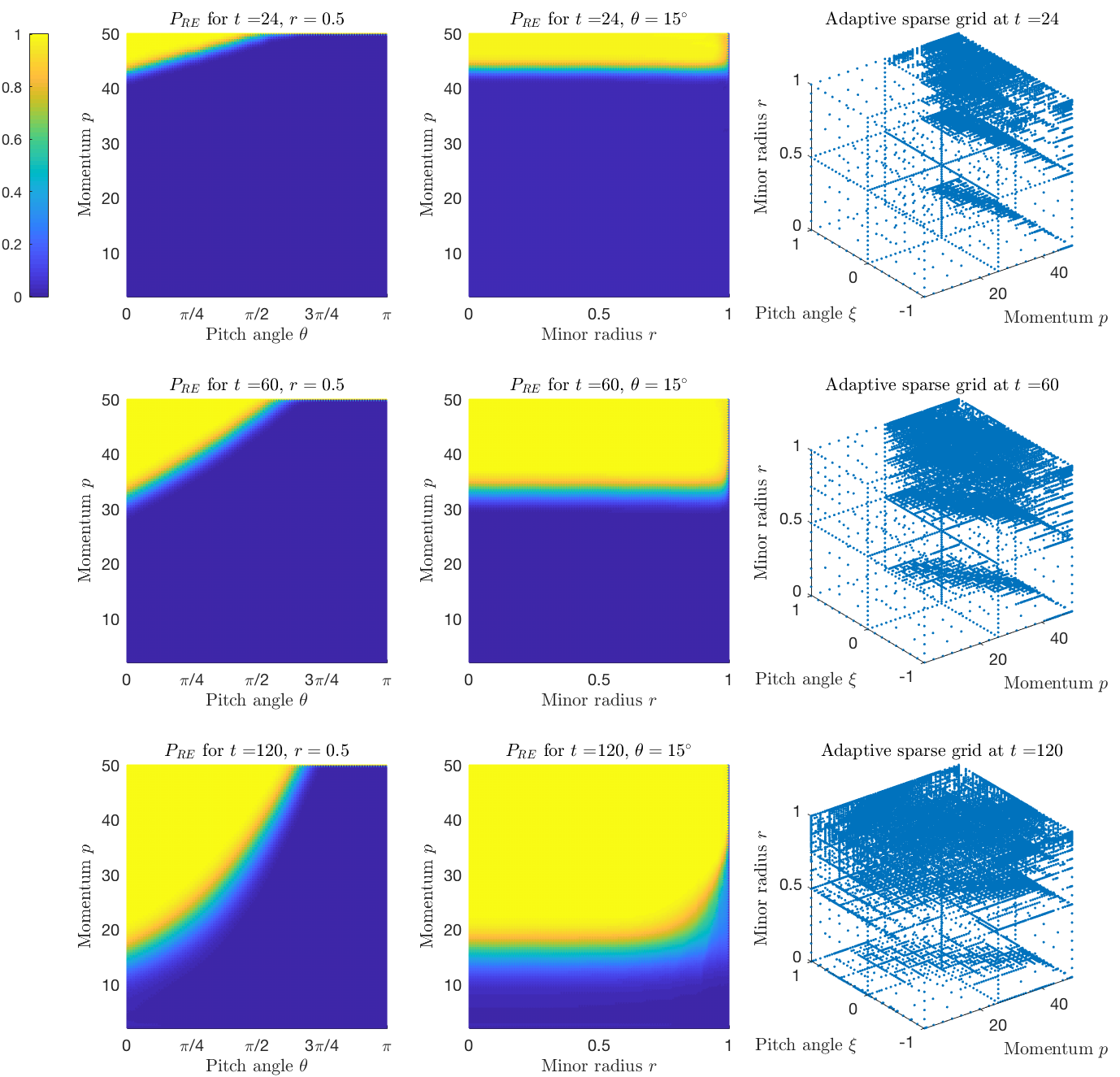}
  \caption{Cross sections of the runaway probability $P_{\rm RE}$ as well as the corresponding adaptive sparse grids at three instants of time $t = 24, 60$ and 120.}
  \label{fig:ex2}
\end{figure}
The standard refinement approach is to construct an initial grid using all points up to some coarse level,
then consider the hierarchical surplus coefficients, e.g., the coefficients of the basis functions,
{\color{black} which are estimates of the local approximation error in the neighborhood of the associated nodes.
The coarse grid is refined by adding the children of nodes with large coefficients ignoring all other points.
Such refinement process is repeated until all coefficients fall below some desired tolerance.}
However, the standard refinement process may stagnate when dealing with functions with localized sharp behavior 
which results in non-monotonic decay of the coefficients (in the pre-asymptotic regime).
In such scenario, a node located in the sharp region could have parent nodes with small surpluses, such that a necessary refinement
will be missed in the local sharp region.
Even if descendants of the node converge in the sharp region (following paths through other parents),
the children have restricted support such that they cannot compensate for the missing parent.
A common remedy for this problem is to recursively add all parents of all nodes,
but this not desirable as it includes many nodes with small coefficients 
which would have been ignored in the classic refinement.
{\color{black} Therefore, we utilized a more flexible refinement procedure that considers both parents and children of nodes with large coefficients, so as to  improve stability and avoid oversampling. Specifically, for each node on the current sparse grid, we first build a set that include both parents and children of the node. Then, we add the children nodes to the sparse grid only if all the parent nodes are already included in the current grid. }
The parents selective refinement procedure is described in details in \cite{stoyanov2018adaptive}
and it is implemented in the TASMANIAN open source library\cite{stoyanov2015tasmanian, doecode_6305}.
\begin{figure}[h!]
\center
  \includegraphics[scale = 0.2]{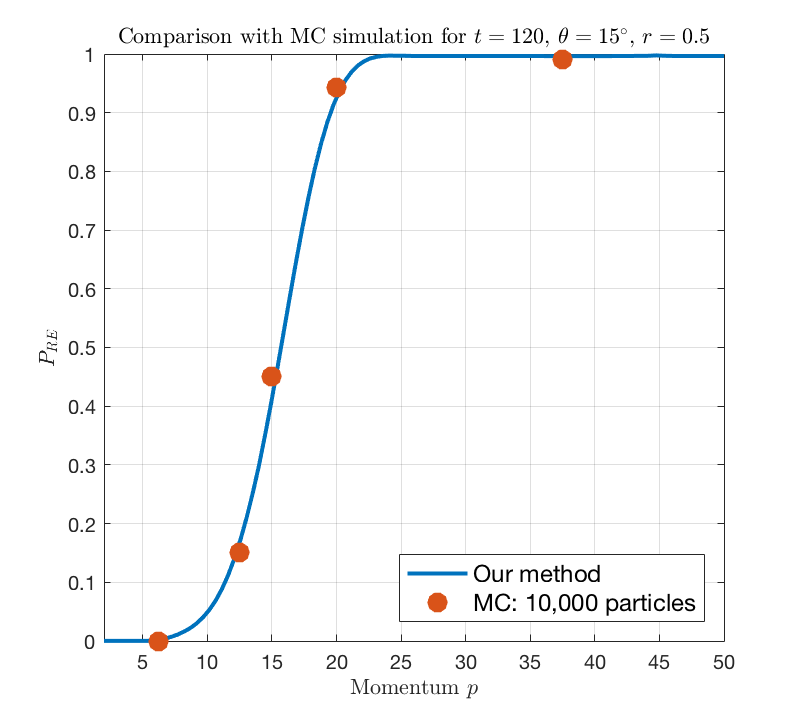}
  \caption{Comparison between the our approach and the direct MC for pitch angle $\theta = 15^{\circ}$ and minor radius $r = 0.5$.}
  \label{fig:ex3}
\end{figure}

The evolution of the runaway probability $P_{\rm RE}$ as well as the corresponding adaptive sparse grids are shown in Fig \ref{fig:ex2}. 
The runaway boundary is at $p = p_{\max} = 50$. 
The main reason of an electron running away is the electric field acceleration, i.e., the   
term $E \xi\,$ in the drift of the momentum dynamics. The factor $\xi = \cos(\theta)$ in $E \xi\,$ determines that the electrons with small pitch angles will runaway sooner than the electrons with large pitch angles, which is consistent with the simulation results in Fig \ref{fig:ex2}. There are two sharp transition layers in this simulation, i.e., the transition between the runaway and the non-runaway regions, and the boundary layer around $r = 1$ due to small diffusion effect in the minor radius direction. In our simulation, we used the 6-level sparse grid as the initial grid and gradually refine it with the tolerance being $0.001$. As expected, the adaptive refinement accurately captured the irregular behaviors. In addition, since the analytical expression of $P_{\rm RE}$ is unknown, we tested the accuracy of our approach by comparing with the direct Monte Carlo method for computing $P_{\rm RE}$ at a few locations in the phase space, and the result is shown in Fig \ref{fig:ex3}. {\color{black} We can see that the RE probability obtained by our approach is consistent with the MC simulations with 10,000 particles, which numerically demonstrate the accuracy of our method. Another observation is that the MC method can only compute $P_{\rm RE}$ at one location in the phase space at a time, such that $P_{\rm RE}$ at the five locations shown in Fig \ref{fig:ex3} requires five repeat simulations of 10,000 particles with different initialization. In comparison, our method can compute $P_{\rm RE}$ at all locations by solving the adjoint equation only once.}

%
%

\section{Concluding remarks} \label{sec:con}
We proposed a sparse-grid probabilistic scheme for the accurate and efficient computation of the time-dependent probability of runaway. The method is based on the direct numerical solution of the Feynman-Kac formula. At each time step
the algorithm reduces to the computation of an integral involving the previously computed probability of runaway and the Gaussian propagator. Sparse-grid interpolation is utilized to recover the runaway probability function as well as evaluate the quadrature points for estimating the conditional expectation in the Feynman-Kac formulation. {\color{black} The integration of sparse grid into the probabilistic scheme provides a fully explicit and stable algorithm to compute the escape probabilities of stochastic dynamics with $\mathcal{O}(\Delta t)$ convergence. Moreover, the adaptive refinement strategy is demonstrated to be effective in capturing the movement of the sharp transition layer of the runaway probability function.} In our future work, we intend to extend our approach to higher dimensional RE problems involving more complicated dynamics. For example, an important RE model to be resolved is to incorporate the relativistic guiding center equations of electron motion into the RE scenario. In this case, the deterministic dynamics of the guiding center motion is six order of magnitudes smaller than the collisional dynamics, which presents significant challenge to the design of numerical schemes.


%
%
%
%


\end{document}